# INFORMATION SOURCES AND ANXIETY AMONG REFUGEES IN KENYA DURING COVID-19

Matthew A. McGee, Pennsylvania State University, mam8223@psu.edu

Carleen Maitland, Pennsylvania State University, cmaitland@ist.psu.edu

Dorothy Njoroge, United States International University-Africa, dwanjiku.n@gmail.com

**Abstract:** In the COVID-19 pandemic, refugees' access to information has become increasingly important given the rapid change in the scientific and public health knowledge-base. However, this access is complicated by social distancing requirements that disrupt traditional in-person communication. Many refugees must then rely on alternative information sources to stay informed. Differences in media types and information sources in turn may be related to anxieties arising from the virus and perceptions of others' adherence to recommended protective behaviors. We examine these relationships with survey data from 1,000 refugees living in both camps and non-camp settings in Kenya. Using logit models, we test relationships between information source and anxiety and the effect of these variables on refugees' expected behaviors of community members. Our primary contributions include the finding that information sources consistently exacerbate (e.g., Facebook) or ameliorate (e.g., news from the internet) different anxieties, or can have mixed effects (e.g., radio). We also find that anxiety and information have significant impacts on refugees' expectations of compliance by others and that, whether between camps or between camps and non-camp locales, findings vary by location. Our results have implications for refugee media and infectious disease anxiety scholarship as well as for managing infectious disease response.

## 1. INTRODUCTION

The COVID-19 pandemic has affected people across the globe, including refugees. It has disrupted traditional inter-personal communication channels, making information on COVID-19 more difficult to obtain. This lack of information may exacerbate refugees' pre-existing anxieties. However, a variety of media, from national and local radio, to mobile-phone-based sources (SMS/WhatsApp) may fill the gap, and help assuage pandemic fears. In general, it has been found that during pandemics, people's adherence to recommended behaviors is influenced by individual factors, such as self-efficacy, and social factors (Farooq et al., 2020), including perceptions of others' adherence (Desclaux et al., 2017). In turn, media and information sources may influence these perceptions (Allington et al., 2020). These media sources may play an even greater role in refugee camps where access to information is more limited due to their typically rural locations and smaller personal networks. We investigate these relationships among refugee populations, which may have locationally determined media and information sources.

To study these relationships, we analyze the COVID-19 related information sources, anxiety sources, and perceptions of other's adherence to virus mitigation strategies reflected in data from the United Nations High Commissioner for Refugees' (UNHCR) Kenya COVID-19 Rapid Response Survey. The data set includes 16 information sources and 10 distinct sources of Covid-19 anxiety. We analyze relationships between information sources and anxiety through logistic regression models for specific locales. Next, using logistic regression, we examine the combined impact of information sources and anxiety on perceptions of others' compliance with recommended behaviors.

The paper is structured as follows. First is the review of the literature on refugees and infectious disease outbreaks, including their media and information environments, disease-related anxiety, and the behavioral intentions and social perceptions related to recommended behaviors. Second is background information on the media and information environment in Kenya as well as a descriptive analysis of the dataset. This is followed by presentation of the findings. Finally, is a discussion of the implications of this study.

## 2. LITERATURE REVIEW

In the following sections we present the aforementioned three-part literature review. Along with describing how media, anxiety, and expectations of others play a role in the normal lives of refugees, we analyze how they may differ under pandemic settings and consider the interactive relationships between them.

### 2.1. Media and Information Environment

Two established media theories, the knowledge gap and media repertoires, inform our analysis of the media environment of refugees. The knowledge gap theory provides the rationale for media use in accessing Covid-19 information, while media repertories emphasizes the multi-media environment of refugees. In the past, the knowledge gap theory has been used to describe the inequality in information acquisition from mass media due to socioeconomic factors (Tichenor, Donohue, & Olien, 1970). However, research on information attainment related to infectious diseases suggests more nuanced analyses are required (Ho, 2012; Bekalu and Eggermont, 2014). Ho (2012) found that the knowledge gap hypothesis was not supported in her study of public knowledge of the H1N1 flu pandemic in Singapore. However, differences were found in the knowledge gap size within her sample based on the type of media used. Bekalu and Eggermont (2014) show that mass media usage widens the regional gap of HIV/AIDS knowledge between urban and rural communities in northwestern Ethiopia. This importance of location in infectious disease knowledge was corroborated by Agegnehu and Tesema (2020) in their study of HIV/AIDS in Ethiopia. The results from these studies show that, while socioeconomic status and media usage are not predictive of infectious disease knowledge, gaps in knowledge of infectious diseases tend to have other underlying influences, like community location and media type.

The media repertoires framework recognizes diverse media use behaviors and underpins our analysis of the relationships between media. Kim (2016), in a study of Korean media use, identified five different media repertoires (e.g., news on traditional media, internet only) and explains these differences with both individual and structural factors. Similarly, Jang and Park (2016) examine complementarity and substitutability between media. Their analysis of Korean users and the role of devices investigates substitutability between paper and computers in consuming news, and computers and telephones in consuming informative content. Peters and Schroder (2018) combine media repertoires with news repertoires, integrating a socio-spatial component. However, the spatial component is on a micro-scale, referring to local situations of media use, such as within a household or on a bus.

Spatial elements are important in the context of refugee communities, as many reside in camps or urban centers. These dense collections of individuals can increase the likelihood of heterogeneity in media sources. Both within camps and between camp-based and urban refugees, differences in wealth, education, literacy, and multilingualism can all affect access to Covid-19 information. As noted by Kumpel (2020), simple heterogeneity in interest levels can affect people's access to news, particularly via social media networks where algorithms use past behaviors to serve up personalized content. This heterogeneity in demographic factors can lead to differences between groups in the usage of media to gain information on and cope with anxiety caused by COVID-19 (Pahayahay and Khalili-Mahani, 2020).

## 2.2. Anxiety

The stresses of forced displacement prior to, during, and post-migration contribute heavily to an increased risk of anxiety among refugees (Hameed et al., 2018). For example, Karenni refugees residing along the Burmese-Thai border see anxiety rates as high as 42% (Vonnahme et al., 2015). Hameed et al. write in their 2018 review article that this increased prevalence of anxiety is due to trauma from violence and political oppression, the resulting lack of basic necessities, and lasting concerns of safety and uncertainty for the future.

Infectious diseases like COVID-19 can exacerbate anxieties by triggering pre-existing traumas, hindering the flow of information, and furthering social isolation. The natural concerns that many people have surrounding illness and death are compounded for refugees, who are in many cases are limited in their ability to check on the health of their friends and family in their home countries (Rees and Fisher, 2020). Anxieties surrounding one's health during an infectious disease outbreak are made worse in refugee camps where medicine and healthcare are already strained due to overcrowding, poor hygiene, and lack of resources (Hermans et al., 2017). Further, the necessary diversion of health resources towards fighting an infectious disease can then limit access to healthcare services for refugees with preexisting mental illnesses (El-Khatib et al., 2020).

An additional source of anxiety for refugees during infectious disease outbreaks is the lack of information due to communication barriers (Bukuluki et al., 2020). Many refugees face communications barriers due to language differences between their home and hosting countries. This barrier can prevent them from receiving public health and educational messages from governmental sources. Language barriers can also expose refugees to public health misinformation from their peers and informal social networks (Bukuluki et al., 2020). This stress-inducing lack of information and confusing misinformation serve as additional sources of anxiety.

COVID-19 has contributed to refugees' anxiety through social distancing policies, which have the potential to exacerbate the already common social isolation of refugees. Social support and connectedness are important for refugee recovery following trauma, but their inability to make new connections due to social distancing inhibits this (Rees and Fisher, 2020). A lack of social support and connectedness, along with financial hardships due to social distancing, can contribute to increased feelings of loneliness and anxiety (El-Khatib et al., 2020). The isolation effects of social distancing may be even more exaggerated for refugees, who tend to be seen as transmitters of the virus due to their migratory status (Bukuluki et al., 2020).

Quarantining policies can be especially traumatic for refugees as well, some of whom have previously experienced forced detainment and restricted freedom of movement (Rees and Fisher, 2020). Properly explained reasoning for the importance of quarantining could alleviate this anxiety, but the aforementioned communication barriers do not always make this possible. This same issue often results in a fear of the uncertainty related to quarantining policies (El-Khatib et al., 2020). In some situations, stringent movement restrictions can even prevent refugees from accessing resources like communal sources of water (Bukuluki et al., 2020). Disruptions to refugees' rights and daily routines like these act as additional sources of anxiety.

## 2.3. Behavioral Intentions and Social Perceptions

Adherence to quarantining and other social distancing policies is influenced in part by social and cultural norms (Webster et al., 2020). During the 2003 SARS outbreak in Toronto, individuals reported that they felt social pressures to adhere to quarantining (Cava et al., 2005). These pressures are dual sided, though, in that when it is known that some individuals have broken quarantine, others are more likely to break protocols as well (Braunack-Mayer et al, 2013). Desclaux et al. (2017) find that cultural pressures to quarantine during the 2014-2016 West Africa Ebola epidemic existed on both the household and the community level. They write that study participants recognized the collective commitment to protecting their community, and that failure to comply was seen as disrespectful.

Along with social perceptions, one of the greatest motivating factors for adherence to virus-mitigating behavioral changes is fear of the virus. In their international sample on COVID-19 behavioral changes, Harper et al. (2020) found that fear of the virus and anxiety related to it were predictors of health-compliant behavioral changes. Other factors like moral or political orientation were not nearly as significant in determining behavioral intentions. Clark et al. (2020) saw similar results in their international study on COVID-19 compliance behaviors. They found that those who wanted to prioritize their own health and believed that behavioral changes would have an effect were the most likely to comply.

Media was also found to influence health-protective behavioral intentions. Specifically, Allington et al. (2020) saw that COVID-19 conspiracy theories resulting from misinformation on social media were a contributing factor to non-compliance with health-protective behaviors. On the other hand, usage of traditional forms of broadcast media had a positive relationship with health-protective behaviors. The authors posit that the reason for these trends observed in their UK sample is that broadcast media is regulated, whereas social media is not.

The above discussion highlights the role of location and media type on infectious disease knowledge of refugees, as well as the types and combinations of media use. It also examines refugee anxiety and the exacerbating role of the pandemic. Of note are the informational aspects, such as the relationship between anxiety and lack of information, as well as the effect of social distancing on information. Finally, the literature review highlights the role of fear of the virus, a form of anxiety, as well as media, on intentions to comply with behavioral recommendations.

While these findings suggest a relationship between information sources, anxiety, and Covid-19 expectations of compliance, comprehensive quantitative evidence is lacking. Specifically, the relationship between information sources and particular sources of Covid-19 anxiety require further investigation. Similarly, while logically media and fear of the virus are likely to impact expectations of others' compliance, little evidence of this relationship among refugees exists. In the following, we examine these associations, specifically investigating:

RQ1: How are refugees' information sources and specific Covid-19 anxieties related? What role does location play?

RQ2: How are refugees' information sources and specific Covid-19 anxieties related to expectations of others' compliance to recommended behaviors? What role does location play?

## 3.  BACKGROUND AND METHODS

Our research is influenced by the Kenyan refugee and media context. UNHCR Kenya (2020) reports the majority of refugees originate from Somalia (54%) and South Sudan (24.6%). Refugees primarily reside in camps, with 44% living in Dadaab, 40% in Kakuma/Kalobeyei, and 16% in urban areas (UNHCR, 2021)

Kenya's media environment is highly-competitive, with one of the most vibrant and well-produced media in the African continent (BBC, 2019). With a literacy rate of 82% (World Bank, 2018), Kenyans consume a wide array of media in English, Kiswahili, and local languages. Currently, Kenya has six dailies, regional publications and various kinds of magazines. While newspapers are mainly concentrated in urban areas, they have a pass along rate of 10-14, greatly increasing their reach (Obuya and Ong'ondo, 2019). Furthermore, digital editions also extend the reach. This is alongside more than seventy television stations and over 160 radio stations according to the Media Council of Kenya website.

Kenya's Internet penetration is one of the highest in Africa at 87.2% (Internet World Stats), occurring primarily through mobile phones (83%) (Namunwa, 2019). However, due to high data costs only 13 million of the 43.3 million Kenyans with access to a phone are active users.

Unsurprisingly, radio remains the most popular medium, followed by television. Media concentration and cross-ownership is an area of concern with 5 privately owned media networks dominating media ownership in the country (Obuya and Ong'ondo, 2019). However, the government-owned media, Kenya Broadcasting Corporation has extensive reach across the country and runs several radio stations and a couple of television outlets.

To study how the consumption of COVID-19 information from these sources impacts refugee anxiety and social perceptions, we utilize public survey data from the World Bank. We first provide descriptive statistics of COVID-19 information sources, anxieties, and social perceptions of others among refugees on a regional level for contextual understanding. After conducting ANOVA and Tukey tests for differences in means among social perceptions, we construct a series of logit models in an attempt to explain why. We first test whether individuals' COVID-19 information sources are predictors for their related anxieties and then for whether their information sources and anxieties predict their perceptions of others adherence to protective guidelines. We chose to use logistic regression models for these tests because of the binary nature of our survey data and the utility of testing significance of both the overall model and the individual predictors as well as the availability of metrics on model fit.

### 3.1. Kenya COVID-19 Rapid Response Survey

Our data comes from a publicly available survey by the World Bank, together with Kenyan National Bureau of Statistics, UNHCR, and the University of California, Berkeley, from May to August of 2020. The group conducted a high frequency phone survey in Kenya to monitor the socioeconomic impact of COVID-19 on households throughout the country. Survey participants came from three different sources: randomly drawn from a subset of the 2015/2016 Kenya Integrated Household Budget Survey, randomly generated phone numbers, and UNHCR's list of registered camp-based and non-camp-based refugees. Limiting the sample to those self-reporting to be refugees produced a sample of 948 respondents.

While the broader survey's primary focus was the economic impact of COVID-19, sections also addressed subjective health and well-being, knowledge of COVID-19, household and social relations, media and information sources used for Covid-19 information, anxieties related to the pandemic, and perceptions of the adherence of other households to recommended behaviors (e.g., social distancing). Questions employed an open response format, where responses were coded by the survey team. For instance, participants were asked: "What sources have provided you with information about Covid-19?" with the choice to name a medium or a source. Results identified 16 Covid-19 information sources/media and 10 Covid-19 anxiety variables.

Our dependent variable 'expectations of others' compliance' was measured as the number of households out of 10. This measure is appropriate due to social desirability concerns biasing responses to direct questions about individuals' intentions to comply.

### 3.2 Descriptive Statistics

Table 1 shows the demographics of the four sample locations (Dadaab, Kakuma, and Kalobeyei camps, and other areas) as well as for the overall sample. While the original survey collected data on refugees' county of residence, the anonymized data set released by UNHCR designates location only as a specific camp or 'other.' While UNHCR often interprets 'other' as 'urban,' since it is strictly unknown, here we maintain the dataset label of 'other.' However, in our analysis, following UNHCR, we often imply 'other' as associated with urban locale. Values represented in parentheses show the proportion relative to the locales' sample size. 'No formal' indicates no education and pre-primary education; 'primary' indicates primary education and vocational schools; 'college' indicates middle level college, university undergraduate, and university postgraduate schooling; and 'other' indicates all other forms of education including Madrassa and Duksi.

| Variable | Dadaab | Kakuma | Kalobeyei | Other | Total |
| --- | --- | --- | --- | --- | --- |
| Population | 218,730 | 196,666 | 38,546 | 80,750 | 494,289 |
| N | 119 (0.0005) | 384 (0.0019) | 223 (0.0058) | 258 (0.0032) | 948 (0.002) |
| Gender | | | | | |
|   Male | 64 (0.54) | 183 (0.53) | 142 (0.64) | 156 (0.60) | 545 (0.57) |
|   Female | 55 (0.46) | 165 (0.47) | 81 (0.36) | 102 (0.40) | 403 (0.43 |
| Age | | | | | |
|   18-23 | 15 (0.13) | 72 (0.21) | 11 (0.14) | 45 (0.17) | 164 (0.17) |
|   23-28 | 22 (0.18) | 65 (0.19) | 32 (0.24) | 53 (0.21) | 193 (0.20) |
|   28-33 | 20 (0.17) | 51 (0.15) | 53 (0.16) | 45 (0.17) | 141 (0.16) |
|   33-38 | 13 (0.11) | 49 (0.14) | 35 (0.20) | 34 (0.13) | 142 (0.15) |
|   38-43 | 9 (0.08) | 42 (0.12) | 46 (0.11) | 25 (0.097) | 100 (0.11) |
|   43-48 | 12 (0.10) | 31 (0.09) | 24 (0.045) | 23 (0.09) | 76 (0.08) |
|   48-53 | 12 (0.10) | 16 (0.05) | 10 (0.04) | 13 (0.05) | 50 (0.05) |
|   53-58 | 7 (0.06) | 12 (0.03) | 9 (0.005) | 9 (0.04) | 29 (0.03) |
|   58-63 | 5 (0.04) | 5 (0.01) | 1 (0.01) | 5 (0.02) | 17 (0.018) |
|   63 | 4 (0.03) | 5 (0.01) | 0 (0.00) | 6 (0.023) | 15 (0.015) |
| Education | | | | | |
|   No formal | 38 (0.32) | 66 (0.19) | 39 (0.17) | 27 (0.10) | 170 (0.18) |
|   Primary | 35 (0.29) | 99 (0.28) | 66 (0.30) | 72 (0.28) | 272 (0.29) |
|   Secondary | 27 (0.22) | 129 (0.37) | 84 (0.38) | 98 (0.38) | 338 (0.35) |
|   College | 4 (0.03) | 28 (0.08) | 23 (0.10) | 47 (0.18) | 102 (0.11) |
|   Other | 15 (0.13) | 26 (0.08) | 11 (0.05) | 14 (0.06) | 66 (0.07) |

**Table 1. Sample Demographics by Location (%)**

In each of the locations the sample consists of more men than women, with the greatest difference being in the Kalobeyei Settlement where 64% of the respondents are male. For privacy, participants ages are partitioned into 5-year intervals. Participants ages ranged from 18 to 63 and above, with over 50% falling in the age range of 18 to 38, both in the total sample and in each location. Some variation exists between locations with regards to educational attainment, but in all cases the majority of individuals have attained either primary or secondary education. The Dadaab camp has lower educational attainment rates, but higher forms of other education including madrassa. In the non-camp locations, refugees have higher educational attainment levels.

In Tables 2 and 3 below, values represent the number of participants in each location mentioning the information source or anxiety respectively, with the relative proportion to the sample in parentheses. Since these values come from multiple response questions, proportions will not sum to 1 for any location.

As indicated in Table 2, information sources ranged from traditional mass media (newspapers, radio, TV) and newer forms (internet news, Facebook) as well as traditional interpersonal sources such as friends and family. In between these broadcast and inter-personal modes, are a variety that could enable sharing of information with small groups or communities (e.g., SMS/WhatsApp) and focus on the source rather than the medium (religious leaders, local medical professionals, etc.). In the combined sample, the most common information sources, in order, are national radio, TV, news on the internet, international or government agencies, and WhatsApp or SMS. It's interesting to note that four of the top five identified are in fact the medium rather than the source.

| Variable | Dadaab | Kakuma | Kalobeyei | Other | Total |
|---|---|---|---|---|---|
| TV | 15 (0.13) | 72 (0.21) | 39 (0.18) | 153 (0.59) | 279 (0.29) |
| Newspaper | 2 (0.20) | 18 (0.052) | 10 (0.045) | 20 (0.078) | 50 (0.053) |
| National Radio | 25 (0.20) | 99 (0.28) | 84 (0.38) | 112 (0.43) | 320 (0.34) |
| Local Radio | 63 (0.53) | 50 (0.14) | 16 (0.072) | 26 (0.10) | 155 (0.16) |
| Billboards or Posters | 2 (0.017) | 11 (0.032) | 10 (0.045) | 5 (0.019) | 28 (0.03) |
| International or Government Agencies | 33 (0.28) | 104 (0.30) | 72 (0.32) | 52 (0.20) | 261 (0.28) |
| NGOs or CBOs | 12 (0.10) | 74 (0.21) | 62 (0.28) | 6 (0.023) | 154 (0.16) |
| News on the Internet | 17 (0.14) | 84 (0.24) | 50 (0.22) | 115 (0.45) | 266 (0.28) |
| Facebook | 4 (0.034) | 45 (0.13) | 11 (0.049) | 33 (0.13) | 93 (0.098) |
| Twitter | 2 (0.017) | 12 (0.034) | 5 (0.022) | 7 (0.027) | 26 (0.027) |
| WhatsApp or SMS | 10 (0.084) | 75 (0.22) | 42 (0.19) | 79 (0.30) | 206 (0.22) |
| Friends, family, or colleagues | 32 (0.27) | 69 (0.20) | 54 (0.24) | 55 (0.21) | 210 (0.22) |
| Local medical professional | 1 (0.008) | 16 (0.046) | 20 (0.09) | 7 (0.027) | 44 (0.046) |
| Other medical professional | 1 (0.008) | 13 (0.037) | 5 (0.022) | 4 (0.016) | 23 (0.024) |
| Political leaders | 1 (0.008) | 3 (0.009) | 1 (0.004) | 1 (0.004) | 6 (0.006) |
| Religious leaders | 7 (0.059) | 15 (0.043) | 12 (0.054) | 3 (0.012) | 37 (0.039) |

**Table 2. Information Variables by Location (%)**

For most information sources, significant variation exists between locations. In 'other' locations the traditional mass media of TV and national radio were the most popular. The most evident location-specific differentiation is the usage of local radio in Dadaab, where over 50% of survey participants reported using the medium as a COVID-19 information source, while less than 15% of participants in all other locations did. This is most likely due to the local Somali-language radio station built and operated in Dadaab. Notable differences also exist between camps in use of news on the internet and WhatsApp/SMS, which are hardly mentioned in Dadaab but account for roughly 20% of mentions in the other two camps.

In a similar fashion, the frequencies of anxiety sources for each location and the total sample are available in Table 3. While a significant factor structure was not identified, subjects identified anxiety sources that roughly fall into three categories: individual and community health, social welfare systems, and broader economic concerns and uncertainty. Anxiety related to the two specific variables death and infection are by far the greatest reported, both overall and consistently in each location. Over 65% of all participants cite them as a source of anxiety.

Like the information source variables, some variation exists between locations for each anxiety source, however it is less pronounced. The greatest difference between locations is in anxiety related to food security, with the Dadaab camp again being the stand out case, where 32% of surveyed individuals reported this anxiety, as opposed to the 20% average in the other three locations. Also, with the lowest proportions of anxieties concerning food, healthcare, and education, refugees living outside of camps have the highest proportion of reported anxieties related to employment and the economy, and 14% cite 'uncertain ends' as an anxiety source.

| Anxiety Source | Dadaab | Kakuma | Kalobeyei | Other | Total |
|---|---|---|---|---|---|
| Infection | 81 (0.68) | 247 (0.64) | 164 (0.74) | 177 (0.69) | 669 (0.71) |
| Death | 70 (0.59) | 232 (0.60) | 147 (0.66) | 171 (0.66) | 620 (0.65) |
| Infecting Others | 42 (0.35) | 129 (0.34) | 67 (0.30) | 58 (0.13) | 296 (0.31) |
| Employment | 7 (0.059) | 48 (0.125) | 24 (0.11) | 50 (0.19) | 129 (0.14) |
| Economy and Mobility | 11 (0.092) | 46 (0.12) | 29 (0.13) | 47 (0.18) | 133 (0.14) |
| Education | 10 (0.084) | 57 (0.15) | 26 (0.12) | 18 (0.07) | 111 (0.12) |
| Healthcare | 24 (0.20) | 71 (0.19) | 26 (0.12) | 18 (0.07) | 139 (0.15) |
| Food Insecurity | 38 (0.32) | 99 (0.26) | 41 (0.18) | 43 (0.17) | 221 (0.23) |
| Uncertain End | 2 (0.017) | 29 (0.076) | 14 (0.063) | 37 (0.14 | 82 (0.086) |
| Other | 1 (0.008) | 9 (0.023) | 8 (0.036) | 3 (0.012) | 21 (0.022) |

**Table 3. Anxiety Variables by Location (%)**

Next, is the data reflecting refugees' perceptions of the number of households following COVID-19 preventative measures. Summary data are depicted in Figure 1 with each panel representing the location in its title. Most respondents believe that all other households (10/10) in their community are adhering to preventative measures. Variation does exist, however, with 21% believing that 5 or fewer households out of 10 are following these practices.

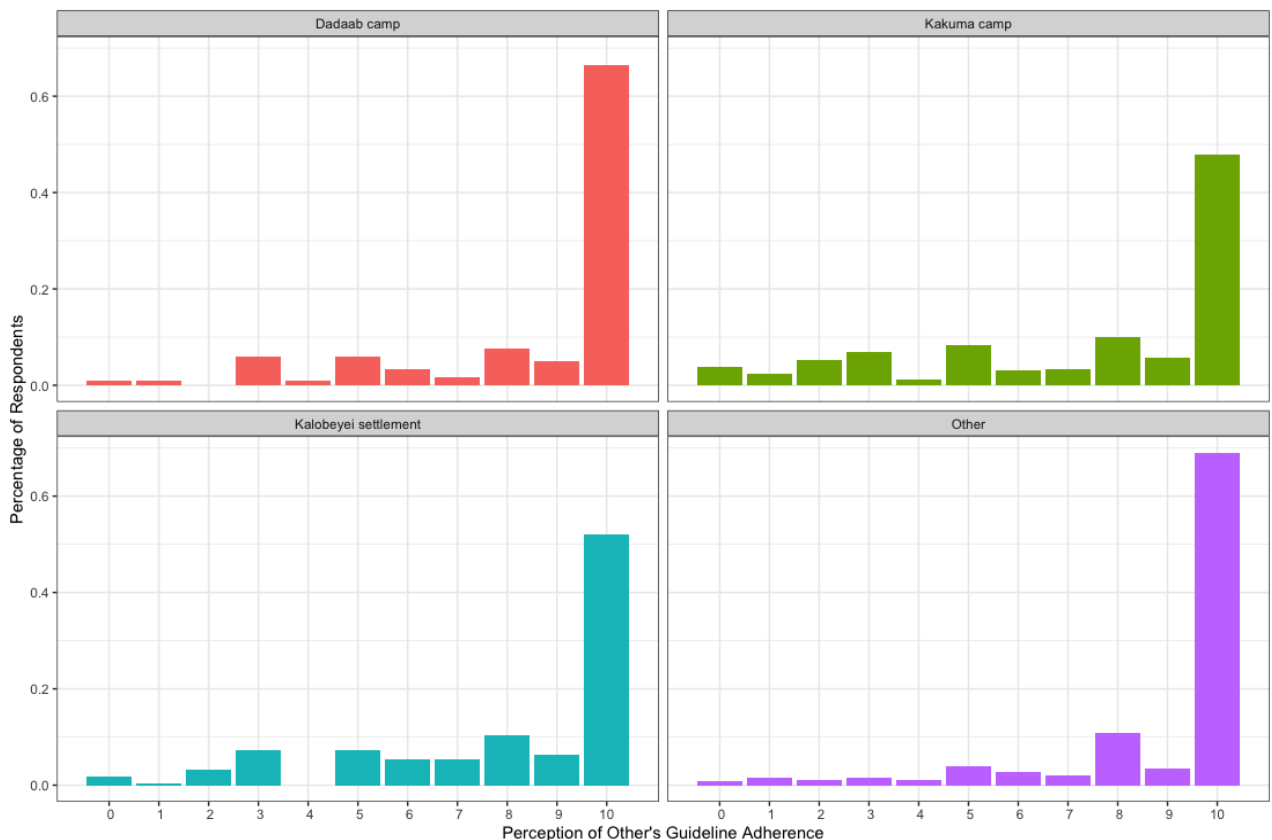

**Figure 1. Perception of Adherence**

Overall, on average, respondents believed roughly 8 out of 10 households in their community were compliant, but a one-way analysis of variance test shows statistically significant differences across locations. The location with the highest average is 'other' locations (8.80), followed by Dadaab Camp (8.68), then Kalobeyei (8.00) and Kakuma (7.49), in last. A follow up Tukey multiple pairwise-comparison confirms a statistically significant difference between the means of Kakuma and Dadaab, Kakuma and other locations, and Kalobeyei and other locations. These analyses are shown in full in Tables 4 and 5, with ** indicating p<0.05 and *** indicating p<0.01. With this knowledge that perceptions of others vary by location, we have reason to investigate if this is due to differing information sources and related anxieties.

|  | Df | Sum Sq | Mean Sq | F value | Pr(>F) |
|---|---|---|---|---|---|
| Location | 3 | 315 | 105.16 | 13.99 | 6.34e-9*** |
| Residuals | 928 | 6977 | 7.52 | | |

**Table 4. One-way ANOVA Table for Perception of Adherence**

|  | Difference | Lower | Upper | Adjusted p |
|---|---|---|---|---|
| Kakuma-Dadaab | -1.1767 | -1.933 | -0.4206 | 0.0003887*** |
| Kalobeyei-Dadaab | -0.6707 | -1.478 | 0.1362 | 0.1415 |
| Other-Dadaab | 0.2023 | -0.5867 | 0.9912 | 0.9121 |
| Kalobeyei-Kakuma | 0.5060 | -0.1034 | 1.115 | 0.1422 |
| Other-Kakuma | 1.379 | 0.7934 | 1.865 | <0.0001*** |
| Other-Kalobeyei | 0.8729 | 0.2232 | 1.523 | 0.003190** |

**Table 5. Tukey HSD for Perception of Adherence**

## 4. FINDINGS

Next, we examine our research questions using logistic regression models. For a first look at the effect of information sources on anxieties, a logistic regression model for each anxiety source was created. In each model (see Table 6), the anxiety source serves as the dependent variable and the information sources act as independent variables. Each column in the table represents a logistic regression model for the corresponding anxiety source. Values in each row are the model coefficients. The first row displays the Nagelkerke pseudo-$R^2$ value for each model. Each of these models is significant, but adjusted $R^2$ values range from 0.031 to 0.369.

Facebook is the most frequently significant variable (8 out of 9 models) and National Radio and NGOs or CBOs follow closely behind (7 out of 9 models each). Some media source variables tend to have similar coefficient signs across all anxiety models, suggesting these sources systematically exacerbate or ameliorate anxiety. Others have mixed effects.

Assessing information sources and numbers of different anxiety sources affected, the 'consistently exacerbates anxiety' category includes TV (3), newspapers (2), and Facebook (8). Conversely, the information sources in the 'consistently ameliorates anxiety' category are 'News on the Internet' (6 significant anxiety sources), WhatsApp/SMS (3), and 'Friends, family, or colleagues' (5). In the mixed-effects category (negative/positive effects on anxiety sources) are national radio (4/3), local radio (4/2), international/government agencies (2/2), and NGOs/CBOs (6/1). Also, no anxiety source had a consistent relationship with each information source. Instead, it appears that each anxiety source had both positive and negative significant relationships among the information sources. This further emphasizes the influence of the information source on anxiety.

Also of note are the models with the lowest and highest adjusted $R^2$ values. The models predicting anxiety related to getting infected and dying have the lowest adjusted $R^2$ values, suggesting that consumption of Covid-19 information does not increase refugee anxiety related to personal wellbeing. However, the three models with the highest adjusted $R^2$ values are those predicting anxiety related to healthcare, food, and infecting others. In contrast to the previous models, Covid-19 information consumption plays a relatively more influential role on refugees' anxiety related to communal wellbeing.

| | Infection | Death | Infecting Others | Employment | Economy | Food | Healthcare | Education | Uncertain End |
|---|---|---|---|---|---|---|---|---|---|
| **Adjusted $R^2$ (Nagelkerke)** | 0.031** | 0.049*** | 0.26*** | 0.117*** | 0.116*** | 0.253*** | 0.369*** | 0.139*** | 0.107*** |
| **TV** | 0.23 | 0.37** | 0.22 | 0.88*** | 0.52** | 0.11 | 0.21 | 0.30 | 0.30 |
| **Newspaper** | -0.06 | 0.20 | 0.78** | 0.59 | -0.29 | 0.52 | 0.86* | 0.17 | -0.076 |
| **National Radio** | 0.35** | 0.38** | -0.85*** | 0.15 | -0.063 | -1.02*** | -1.62*** | -0.76*** | 0.49** |
| **Local Radio** | 0.003 | 0.32 | 0.62*** | -1.14*** | -1.17*** | 0.48** | -0.004 | -1.22*** | -1.28** |
| **Int or Gov Agencies** | 0.54*** | 0.28* | 0.02 | 0.17 | 0.30 | -0.62*** | -0.55** | 0.057 | 0.39 |
| **NGOs or CBOs** | -0.07 | 0.55*** | 0.04 | -0.86** | -1.12*** | -0.56** | -1.58*** | -1.14*** | -1.19** |
| **News on Internet** | -0.35* | -0.13 | -1.38*** | -0.73*** | -0.091 | -1.54*** | -2.11*** | -0.592** | 0.31 |
| **WhatsApp or SMS** | -0.10 | -0.15 | -0.73*** | 0.028 | -0.31 | -0.55** | -1.13*** | -0.38 | 0.083 |
| **Facebook** | 0.23 | 0.59** | 1.78*** | 0.745** | 1.29*** | 1.92*** | 2.48*** | 1.50*** | 0.92*** |
| **Friends, family or colleagues** | 0.02 | -0.46*** | -1.32*** | -0.081 | -0.42 | -1.23*** | -2.93*** | -0.64** | -0.53 |

**Table 6. Information and Anxiety Logistic Regression Models**

Finally, we conducted logistic regression analyses of expectations of compliance to assess the joint impact of information and anxiety (see Table 7). Variables that had low correlation levels were left out to reduce dimensionality. Preliminary analyses included age and gender, but found they were not significant and did little to improve the overall adjusted $R^2$ value. Education is the only significant demographic variable, and only for refugees living outside of camps.

Information sources' associations with expected behaviors in the combined sample include significant positive effects (national and local radio), as well as significant negative effects (Facebook and newspapers). In contrast, the location-specific effects of information sources were difficult to discern. Between camp comparisons are hindered by differences in camps' information sources. Nevertheless, in the non-camp ('other') locations, TV has a significant positive effect on expectations. In both Dadaab and Kakuma, local radio had a significant positive effect. Hence, the consistent and mixed relationships between information and anxiety discussed above are not evident for information and expectations.

Similarly, the relationship between anxiety and expectations of compliance are mixed. In the combined sample, worries related to infection, infecting others, and education are all negatively related to expectations, whereas fears over the economy had a positive relationship. Yet in Dadaab and Kakuma, anxiety related to death had a positive relationship with expectations. Finally, for those outside camps, anxiety had mixed effects. Similar to the combined sample, worries over infecting others and education had a negative impact on expectations, while fears about the economy had a positive relationship.

Assessing the models and predictors overall, each logistic regression model is significant under Chi-square goodness of fit tests and each has an adjusted $R^2$ value of 0.38 or greater. Across locations, the model's explanatory power is greatest for Dadaab and least effective for non-camp locations. Systematic assessment of information and anxiety effects show some consistency in effect. For example, local radio was a significant predictor in Dadaab and Kakuma. Similarly, death anxiety had a positive effect on expectations in both Dadaab and Kakuma. Consequently, it can be concluded that while information sources and anxieties have differing effects, both positive and negative, and their significance varies by location, consistent effects do exist.

| | All | Dadaab | Kakuma | Kalobeyei | Other |
|---|---|---|---|---|---|
| **Adjusted $R^2$ (Nagelkerke)** | 0.470 | 0.685 | 0.534 | 0.483 | 0.380 |
| **No formal schooling and pre-primary** | - | - | - | - | 1.968 (0.0786*) |
| **Primary and vocational Schooling** | - | - | - | - | 2.225 (0.0166**) |
| **Secondary Schooling** | - | - | - | - | 2.5064 (0.0083**) |
| **College, university and post-graduate** | - | - | - | - | 1.736 (0.0743*) |
| **Infecting Others Anxiety** | -0.7434 (0.0119**) | - | - | - | -1.703 (0.0136**) |
| **Education Anxiety** | -0.8796 (0.0971*) | -17.87 (0.9300) | -0.2292 (0.7445) | -1.373 (0.2805) | -2.359 (0.0301**) |
| **Healthcare Anxiety** | 0.3703 (0.7425) | - | 0.1240 (0.9163) | 7.068 (0.9563) | 0.9573 (0.9964) |
| **Death Anxiety** | -0.2698 (0.4846) | 3.231 (0.0667*) | 0.5760 (0.0979*) | - | -1.1819 (0.1612) |
| **Food Anxiety** | 0.1215 (0.7438) | -2.405 (0.1946) | -0.6700 (0.1603) | 0.7188 (0.5217) | 0.6260 (0.4258) |
| **Economy Anxiety** | 1.760 (0.0282**) | -0.8711 (0.9866) | 0.9490 (0.4042) | 7.463 (0.8506) | 1.909 (0.0689*) |
| **Employment Anxiety** | - | -4.319 (0.1979) | - | - | - |
| **Infection Anxiety** | -0.5846 (0.0683*) | - | - | - | -1.106 (0.1547) |
| **National Radio** | 0.4101 (0.0806*) | 1.628 (0.1662) | 0.7428 (0.0738*) | 0.7685 (0.1336) | - |
| **Local Radio** | 3.366 (0.0001***) | 3.655 (0.0043***) | 2.673 (0.0001***) | - | 9.779 (0.8264) |
| **TV** | - | - | - | - | 1.026 (0.0418**) |
| **News on the Internet** | 0.4818 (0.1366) | 7.045 (0.7645) | 0.8378 (0.0465**) | 0.1533 (0.8289) | 0.3095 (0.6254) |
| **Facebook** | -0.6988 (0.0835*) | - | - | - | - |
| **Newspaper** | -1.131 (0.0042***) | - | -2.576 (0.0004***) | - | - |
| **WhatsApp or SMS** | -0.4983 (0.1337) | -2.063 (0.1740) | - | -1.3917 (0.0177**) | -0.0352 (0.9658) |
| **Healthcare * Food** | -3.205 (0.0091***) | - | -2.973 (0.0309**) | -8.978 (0.9445) | -3.7561 (0.9860) |
| **Education * Economy** | -2.407 (0.0179**) | 14.30 (0.9458) | -2.786 (0.0779*) | -7.5317 (0.8494) | - |
| **Death * Infection** | 1.242 (0.0079***) | - | - | - | 2.612 (0.0132**) |
| **News on the Internet * WhatsApp or SMS** | 1.0826 (0.0536*) | -1.113 (0.9632) | - | 0.8767 (0.4144) | 0.9235 (0.4359) |
| **Local Radio * News on the Internet** | - | -0.8860 (0.9922) | - | - | |
| **Food * Local Radio** | - | - | - | - | 2.624 (0.9644) |

**Table 7. Perception of Adherence Logistic Regression Models (p value)**

## 5. Discussion and Conclusion

Generally, this research extends knowledge of the impact of information sources on refugees' infectious disease anxiety and perceptions of others, taking into account location. Specifically, the study examined the relationships between unique information sources and sources of Covid-19 anxiety, and then the relationship between both information source and anxiety with behavioral expectations of others. These findings contribute further evidence of the role of location on type of media use in infectious disease information dissemination. Following and expanding on the work of Bekalu and Eggermont (2014) and Agegnehu and Tesema (2020) in Ethiopia, our study finds locational differences in media types used in Kenya. In our sample, the traditional mass media of TV and national radio are more commonly used by refugees in non-camp locales, as compared to their camp-based counterparts. However, we also demonstrate differences between rural areas, highlighting the place-based nature of some media types, complementing the micro-scale analysis of Peters and Schroder's (2018) media repertoires research. It is worth noting the population density differences between typical rural communities and refugee camps. While the camps in this study were rurally located and had a similar lack of nearby infrastructure and resources, their population density provides increased potential for social interaction not typically seen in rural communities.

Overall, this research finds refugees' anxiety sources fall into three general categories: individual and community health, social welfare systems, and broader economic concerns and uncertainty. Between camps and other locales, the nature of anxieties differed. In camps, COVID-19 anxieties were related to community health and social welfare, such as food security and healthcare, while refugees living outside of camps were more concerned with broader economic trends. Further, between-camp comparisons highlighted differences, such as the concern over food security in Dadaab, which is less evident in Kakuma and Kalobeyei. These findings extend the research on refugees' anxiety related to infectious disease (Hermans et al., 2017; El-Khatib et al., 2020) by analyzing locational differences in the sources of anxiety.

Critically, this study provides unique insights into the relationships between anxiety sources and media types through regression models. These models help delineate media's relationship with specific sources of anxiety. Media have the lowest impact on anxiety related to personal wellbeing (getting infected and dying) and the highest impact on communal wellbeing (infecting others, healthcare, and food). The analysis also identified media that consistently ameliorate or exacerbate different anxieties such as Facebook (exacerbating) and news on the Internet (ameliorating).

A further unique contribution is insights into the the joint impact of information sources and anxiety on expectations of others' adherence to recommended Covid-19 behaviors. As media are expected to play a role in shaping individuals' perceptions of others' adherence to recommended behaviors (Allington et al., 2020), such expectations can, in turn, influence individual compliance. Of particular note in the analysis is the systematic lack of effect of age and gender, and only a location-specific effect of education. These results deviate from findings related to the role of demographics in general in refugee media use (Pahayahay and Khalili-Mahani, 2020), and gender in particular as relates to refugees' mobile phone use (Canevez et al. 2021). The limited effect of demographics suggests programs promoting compliance can have broad but locationally-specific effects.

This combined analysis builds bridges between the disparate media effects and anxiety research in the specific domain of infectious disease behavioral compliance. Whereas Harper et al. (2020) and Clark et al. (2020) find fear of the virus promotes compliance, Allington et al. (2020) point to media. The results showing fear of death as related to expectations of others in Dadaab and Kakuma lend further support to the notion that fear of the virus is associated with compliance. In addition is the positive relationship between fear of economic impact and compliance expectations. Conversely, this work identifies anxieties that may hinder compliance, namely fear of infecting others and anxiety about education. Similarly, these findings of the positive relationship between national and local radio as well as TV on perceptions of compliance lend further support to Allington et al.'s (2020) findings, but add the caveat that in Kenyan context, these results will vary by locale.

As such, a key contribution of our overall study is the importance of location in not only assessing anxiety and sources of information but on crafting effective responses to reduce anxieties and promote behavioral compliance. This study has highlighted the locational differences not only between urban and rural locales, but between different rural locales – a finding that may be related to unique sociocultural context of refugee camps, which in the Kenyan context tends to host refugees from different nations. Nevertheless, we do find some consistency in effects of information sources on various anxieties, with some that consistently exacerbate (e.g., Facebook) others ameliorate (e.g., news from the internet), while still others having mixed effects (e.g., radio).

 In summary, this study is the first of its kind to examine the relationship between Covid-19 media use and anxiety, with further implications for community expectations concerning compliance with recommended behaviors. It demonstrates significant differences in Covid-19 media use between camps and other locations, as well as between camps themselves. Locational differences in sources of anxiety were also found, although less notable than those in media use. These findings have implications for scholarship on infectious disease management in refugee hosting nations, as well as practical implications for designing information campaigns to reduce anxiety and encourage compliance with recommended behaviors.